\documentstyle[aps,twocolumn,psfig]{revtex}
\begin{document}

\draft

\title{Crossover to Self-Organized Criticality
in an Inertial Sandpile Model}

\author{D.A.Head\cite{em1} and G.J.Rodgers\cite{em2}}
\address{Department of Physics, Brunel University,
Uxbridge, Middlesex, UB8 3PH, UK}

\date{\today}

\maketitle

\begin{abstract}
We introduce a one-dimensional sandpile model which
incorporates particle inertia.
The inertial dynamics are governed by a new
parameter which, as it passes though a threshold value,
alters the toppling dynamics in such a way that the system no
longer evolves to a self-organized critical state. A range of
mean-field theories based on a kinetic equation approach is
presented which confirm the numerical findings. We conclude by
considering the physical applications of this model,
particularly with reference to recent experimental results.
\end{abstract}


\section{Introduction}

The concept of {\em self-organized criticality} (SOC)
was introduced by Bak, Tang and Wiesenfeld~\cite{BTW}
as a possible explanation for
the common occurrence of scale-invariance in nature.
To demonstrate this behaviour, they introduced the
{\em sandpile model}, a driven dissipative cellular automata
whose dynamics are defined by local interactions.
Despite the short-range dynamics,
the system organizes itself into
a non-equilibrium critical state with no
finite correlation length and hence
no characteristic length scale.
A feature common to all SOC systems is that
the sizes of fluctuations follow power-law distributions,
a direct consequence of the scale-invariance. 
However, comparisons with real systems
have met with only partial success.
Power-laws were observed in a granular mixture
when avalanches were initiated by watering the
pile~\cite{exp2}, but not
in a pile of glass beads that was
gradually tilted~\cite{exp3}.
Adding grains individually to a conical sandpile
only showed power-laws for sufficiently small
piles~\cite{exp1}.
Recently, Frette {\em et.al.}~\cite{nature} performed
experiments on one-dimensional piles of rice and
found power-law behaviour only for grains
with sufficiently large aspect ratio.
In light of these experiments, we believe it would be
informative to construct a sandpile model
with greater physical applicability,
whilst hopefully retaining some of the
interesting dynamics.

One important ingredient missing from the basic
model is inertia.
To the best of our knowledge, there have only
been two attempts to construct an
inertial sandpile model.
Prado and Olami~\cite{Prado} chose to
associate moving particles with a decrease in the
local stability,
and found SOC behaviour only for small systems,
in both one and two dimensions,
thus giving an explanation
for the results in~\cite{exp1}.
Krug, Socolar and Grinstein~\cite{KSG} gave a single
measure of momentum to the entire cluster of
moving particles in a one-dimensional system.
They found that their inertia parameter needed
to be fine-tuned to zero for the system to
become SOC.

In this paper, we consider a new way of
incorporating inertia into the sandpile model.
A full description of the model is given in the
following section, but briefly,
we suppose that moving particles only come to rest
on those sites whose slopes are not too steep,
where the definition of `too steep' is
controlled by a new parameter.
Our main result is the crossover
between SOC and non-SOC
behaviour as this parameter passes through
a threshold value.
The existence and value of this transition point
has been confirmed by extensive numerical
analysis, mean-field theories and qualitative
reasoning.

This paper is arranged as follows.
The new model is defined and some of its immediate
consequences explored in section~\ref{sec:model}.
In section~\ref{sec:results}, the results of
numerical simulations are given for the simplest
non-trivial sandpiles. These results are
confirmed by the mean-field analysis given in
section~\ref{sec:MFT}, where a rate equation approach has
been adopted. Finally, in section~\ref{sec:summ}, we
explain the nature of the transition threshold
and discuss applying the model to real physical
systems.

\section{The Model}
\label{sec:model}

A one-dimensional sandpile is defined by a set
of integer heights $h_{i}$, $i=1\ldots L$, or equivalently by
the local slopes $z_{i}=h_{i}-h_{i+1}$.
The right-hand boundary is taken to be open, $z_{L}=h_{L}$,
whereas
the left-hand boundary is treated as closed, $z_{0}=0$.
Particles are added sequentially to randomly chosen sites,
increasing their height by one unit.
In the limited local sandpile (LLS) model~\cite{first},
a site $i$ becomes unstable when $z_{i}$ becomes greater
than the {\em critical slope parameter}
$z_{c}$. Any such unstable site will {\em topple},
$z_{c}$ particles leave site $i$ and
move to site $i+1$ (or leave the system if $i=L$), so

\begin{eqnarray}
z_{i-1}&\rightarrow &z_{i-1}+z_{c}, \nonumber \\
z_{i}&\rightarrow &z_{i}-2z_{c}, \nonumber \\
z_{i+1}&\rightarrow &z_{i+1}+z_{c},
\end{eqnarray}

\noindent{with equivalent rules for toppling at boundaries.
It is now possible for $z_{i-1}$ and/or $z_{i+1}$ to
become unstable and topple, and an avalanche will begin.
We call the series of the initial topplings at sites
$i-1$, $i-2$, $i-3$ \ldots the {\em back-avalanche}.
A back-avalanche propagates
to the first site $j<i$ with $z_{j}\leq 0$, where such sites
are called {\em troughs}. In terms of the slopes
before particle addition to site $i$, $z_{i-1}=1$ will
also stop a back-avalanche. Some previous
literature~\cite{CCGS} refers to
such instances as {\em slide events}. 
Topplings occur on a timescale much smaller and
separated from that of 
particle addition, in that no more particles are
added to the pile until the avalanche is over and every site
has become stable ($z_{i}\leq z_{c}$ $\forall i$).}

As this model stands, the inertia of the toppling particles
has been ignored, any momentum in the topple
is assumed to be dissipated instantaneously from
the sandpile as soon as the particles have moved.
The previous attempts to incorporate
inertia into the system~\cite{Prado,KSG}
used quantities that evolved throughout the
avalanche and hence introduced
a form of memory into the system.
In this paper, we consider a new set of rules for
toppling that requires just one
extra, time-independent quantity, but nonetheless
intuitively mimics inertia.
We introduce the {\em minimum-slope inertial limited
local sandpile}
(MILLS) model, which has a second critical slope parameter,
the {\em inertial threshold} $z_{in}$. Now when a site $i$
topples, the particles will be deposited on the
first site $j>i$ obeying

\begin{equation}
z_{j}<z_{in},
\end{equation}

\begin{figure} [h]
  \begin{center}
    \leavevmode
      \psfig{figure=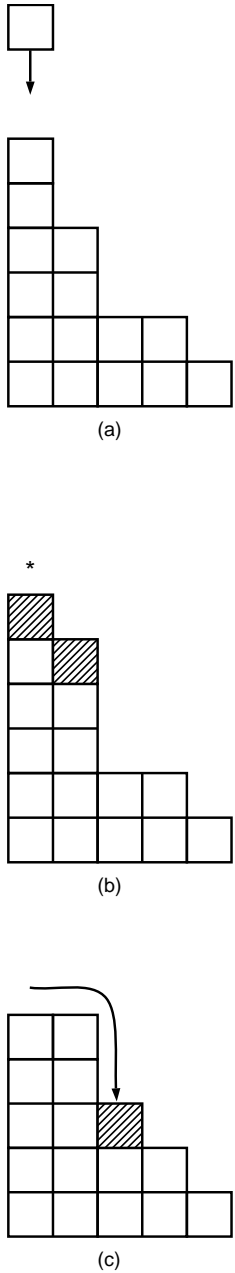,clip=true,width=3cm}
  \end{center}
\caption{An example of the toppling process in the MILLS
model, for $z_{c}=2$ and $z_{in}=2$. (a) Particle added
to a site with critical slope. The site becomes unstable
and topples. (b) The first toppling particle sticks to
the adjacent site, giving it a local slope of 3. The
site marked with an asterisk is still in the process
of toppling. (c) The second
particle slides throgh the site with slope 3 to stop
at the next site. All sites in this region are now stable.}
\label{f:eg}
\end{figure}

\noindent{or (if no such site exists) leave the system.
This calculation is performed right-to-left for each
unstable pile and individually for each of the
$z_{c}$ particles involved in every toppling,
recalculating the slopes as a particle comes
to rest. An example is given in fig.\ref{f:eg} for
$z_{c}=2$ and $z_{in}=3$.
Note that we have implicitly introduced a timeframe
for inertial effects which is much smaller
than that of toppling events. This
separation of timescales means that the time taken for
an unstable site to begin to topple is much
longer than that for the ensuing particle motion itself.}

As a realistic physical system, this model does have some
obvious drawbacks, the most notable being that a particle will
be stopped just as easily if it
has just toppled, or if it has rolled down a
large region of slopes $\geq z_{in}$.
We could of course
make $z_{in}$ dependent on each particle's previous
motion, but it is the
lack of inherent memory that makes much of
the subsequent analysis possible. Furthermore, with this
implementation of inertia the sandpile
takes a microscopically smooth,
realistic shape which further justifies its study.

We can immediately make some general observations about the
MILLS
model. Since the particles toppling from an
unstable site all move at least one step, the slope of the
site to the immediate left will always increase by $z_{c}$
independent of the value of $z_{in}$. Thus, the
back-avalanches in this model are identical
to those in the standard model,
and consequently troughs are still important as bounds
of the left-hand edge of an avalanche. However, troughs no
longer necessarily bound the right-hand edge of an avalanche
as they did in the LLS model. Indeed, no expression
involving a single site can now serve as a
general right-hand bound, so the previous analysis
of the standard model
based on troughs~\cite{KSG,Krug} cannot be extended to
this model.

For $z_{c}=1$, the sandpile soon reaches a
trivial steady state with $z_{i}=1$ $\forall i$ independently
of $z_{in}$. Inertial effects are now
indistinguishable from topplings, added particles
move downslope by either mechanism and leave the system
in an unaltered state. Varying $z_{in}$ will change the time
taken for the particle to reach the right-hand boundary, but
only for $z_{c}\geq 2$ can $z_{in}$ have any influence
on the nature of the critical state.

If $z_{in}=-\infty$, inertial effects will dominate and
all toppling particles will
immediately leave the system, as if the sandpile has become
`infinitely slippery'. This is equivalent to the case
$r=0$ for the model in~\cite{KSG}.
Since sites with negative slope are rare,
$z_{in}\leq 0$ will have a similar effect to
$z_{in}=-\infty$.
For $z_{in}\leq z_{c}$, no particle will
come to rest on a site with critical slope and so
no sites to the right of the initial topple can
become unstable. As a result, there will be
no forward toppling and the avalanche will
consist solely of the back-avalanche.

It may appear that implementing inertia
in this manner could allow for a large cluster
of sliding particles to all come to rest
simultaneously on the same site, say a site $i$
with $z_{i}$ large and negative.
It is easy to prove, however, that at most $z_{c}$
particles can stop on a site during any one
step in the toppling process. To see this, first
realise that a back-avalanche consists of one
topple per avalanche time step, so to get more than
$z_{c}$ particles moving at once we must have
$z_{in}>z_{c}$. Furthermore, a site $k<i$ must topple
and subsequently allow at least one particle to slide
through it, ie.

\begin{equation}
z_{k}-z_{c}\geq z_{in},
\end{equation}

\noindent{so
for this sequence of events to occur requires
$z_{k} > 2z_{c}$.
Although site $k+1$ toppling could contribute
to such a large slope, having $z_{k+1}>z_{c}$
on the previous step would simultaneously
require $z_{k}\leq 0$, except at the left-hand
boundary of the avalanche, as closer analysis
soon reveals. It should now be apparent
that for any site to gain more than $z_{c}$ particles
during any one avalanche step, there must already exist
a site in the system to which this has already happened.
Since particle addition cannot
create such a site, we must conclude that they never occur.
Intuitively, this result implies that inertial effects
serve only to broaden the shape of the avalanche.}

If $z_{in}=\infty$, inertial effects will be so weak that
toppling particles will never move by more than one site,
and we just have the LLS model.
In fact, from the result given in the previous
paragraph, it should be apparent that the maximum slope
that a site can reach is $2z_{c}$, so if 
$z_{in}\geq 2z_{c}$, inertial
effects become redundant
and the MILLS model
reduces to the standard model.

\begin{figure} [h]
  \begin{center}
    \leavevmode
      \psfig{figure=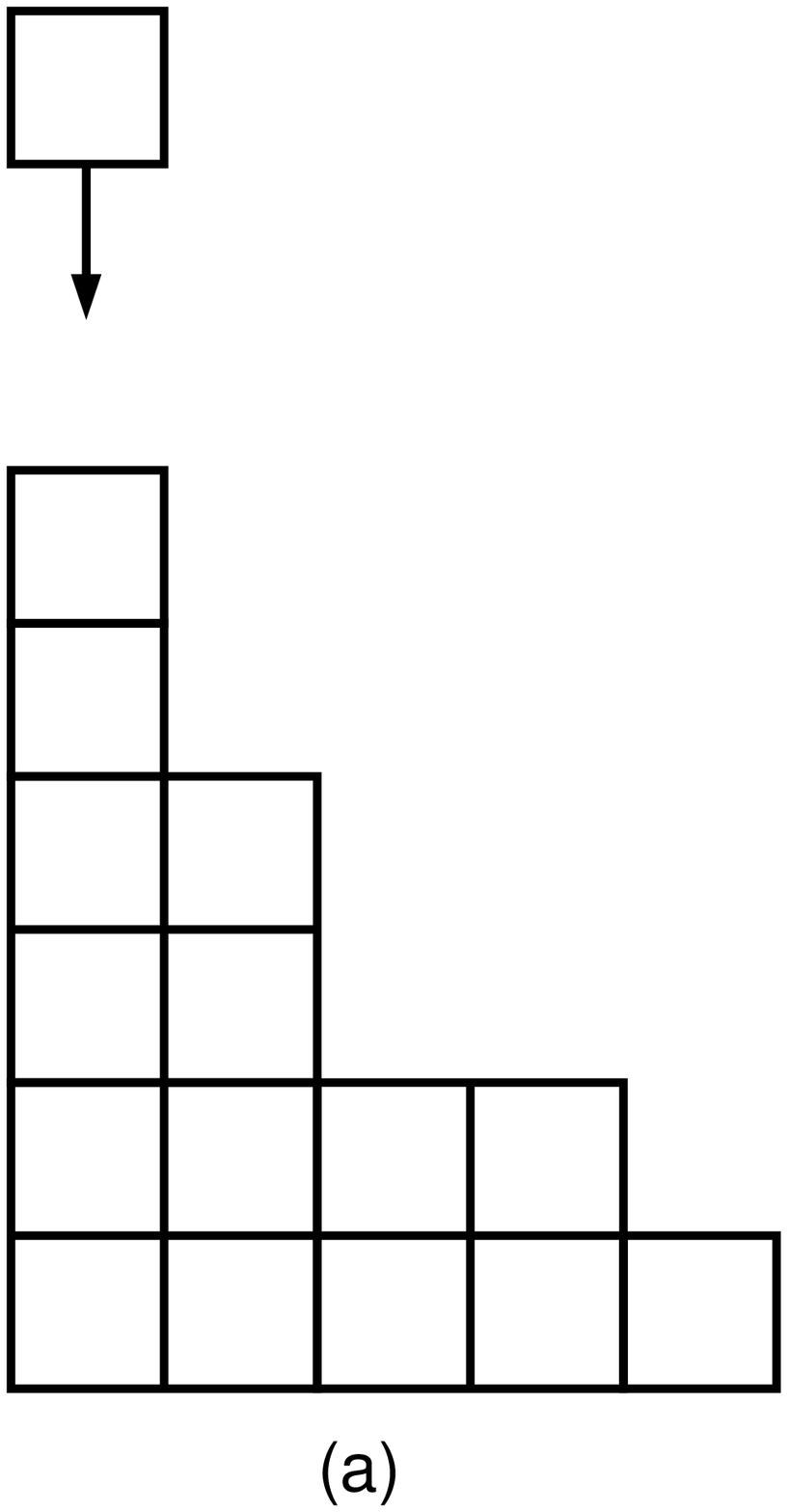,clip=true,width=3cm}
  \end{center}
\end{figure}

\begin{figure} [h]
  \begin{center}
    \leavevmode
      \psfig{figure=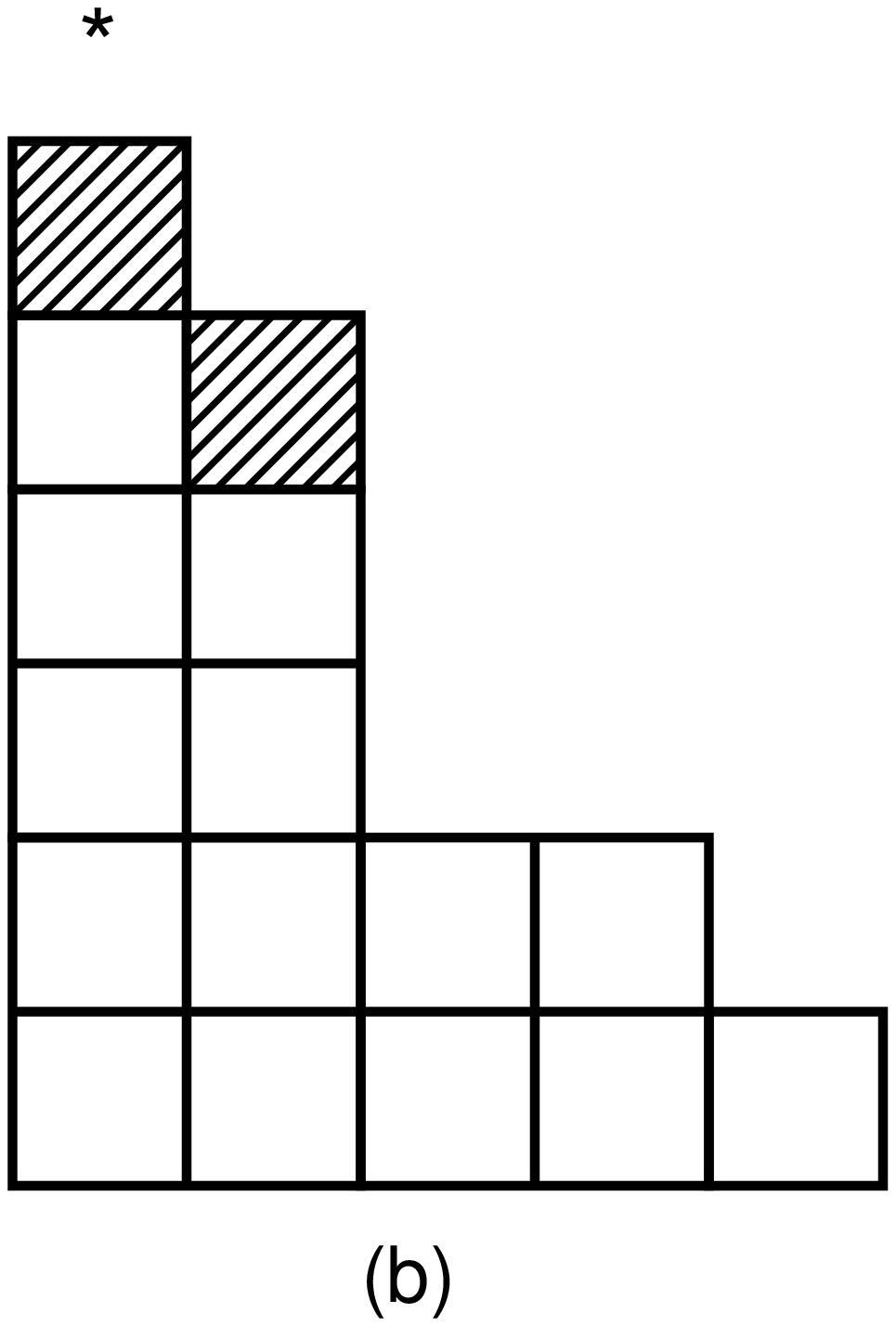,clip=true,width=3cm}
  \end{center}
\end{figure}

In~\cite{nature} it was suggested that it was the
shape of the particles that
determined the nature of the critical pile.
Although any sandpile model is over-simplified
compared to real piles of any substance,
for instance in failing to allow
for the particles to overlap and for ignoring
any variation in particle dimensions, it is still instructive
to consider qualitatively how the two parameters
$z_{c}$ and $z_{in}$ might relate to actual systems.
Suppose the particles all have an aspect ratio $\alpha$.
A vertical stack consisting of rounded particles will
clearly be much less stable than a stack of
flat particles, so $z_{c}$ will decrease as
$\alpha$ decreases. The parameter $z_{in}$
should also decrease with $\alpha$, since
rounder particles will roll more easily and each site
will be narrower and hence easier to traverse.
Indeed, varying $\alpha$ should cause $z_{in}$ to change
faster than $z_{c}$ does, so we can see that the ratio
$z_{in}/z_{c}$ will increase with $\alpha$.
More precise analysis in this manner
is possible but any such accuracy is lost within the
artificial framework common to all sandpile models.

In summary, the MILLS model differs from the LLS model
by the extra parameter $z_{in}$. For $z_{in}\geq 2z_{c}$ we
just get the LLS model, for $z_{in}\leq z_{c}$ the toppling
process consists of just the back-avalanche, and for
$z_{in}\leq 0$ the majority of sliding particles leave the
system immediately after toppling.

\section{Results}
\label{sec:results}

\begin{figure} [h]
  \begin{center}
    \leavevmode
      \psfig{figure=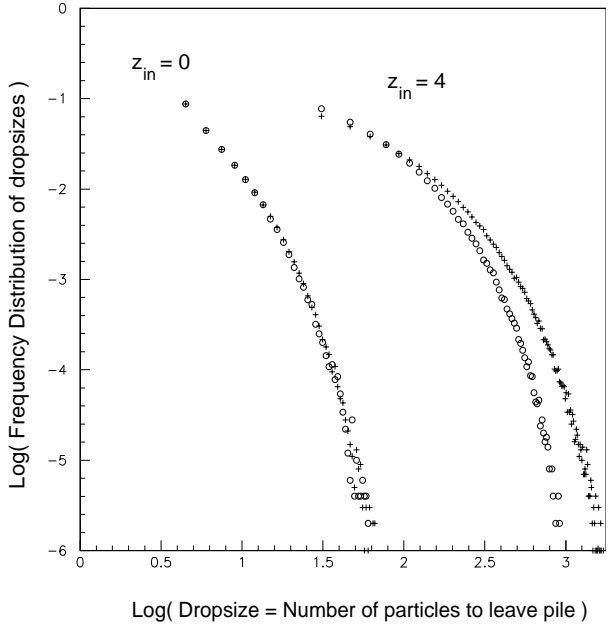,clip=true,width=8cm}
  \end{center}
\caption{Distribution of dropsizes for $z_{in}=4$ and $0$, and
for two different system sizes, with $z_{c}=2$. Circles
correspond to $L=512$, crosses to $L=1024$.}
\label{f:fourandzero}
\end{figure}

Numerical analysis of finite one-dimensional
sandpile models is difficult because
the convergence to the asymptotic SOC
regime is very slow.
Indeed, we are not aware of any simulations sufficiently
large to demonstrate the power-law behaviour expected
in the thermodynamic limit, so an alternative
test for self-organised criticality must be employed.
If the critical state is governed by exponentially
decaying correlations,
then increasing the system size far beyond
the correlation length could not alter the avalanche size.
However, if instead the correlations decay
as power-laws, then the critical state is
scale-invariant and so the largest avalanche sizes
will always vary with the system size.
Two examples of dropsize frequency distributions
are given in fig.\ref{f:fourandzero},
where the dropsize is defined as the number of particles
to leave the pile as the result of a single particle addition.
The distribution broadens for $z_{in}=2z_{c}$ when the
system size is doubled, which is clearly not the case
for $z_{in}=0$, thus the system is not SOC for $z_{in}=0$.
It is obviously important to find 
the value of $z_{in}$ between these two extremes
at which the transition between SOC and
non-SOC behaviour occurs.

The distribution of dropsizes is not always
a useful measure of avalanche size, since for
$z_{in}=z_{c}$ the maximum dropsize is just $z_{c}+1$,
as the following analysis demonstrates. Suppose an avalanche
is started by the addition of a particle onto a site $i$
with critical slope, so that before toppling we have 
$z_{i}=z_{c}+1$ and $z_{i-1}\leq z_{c}-1$. When site
$i$ topples, $z_{i}$ will decrease by at least
$z_{c}$, possibly more if any particles stick to $i+1$,
so now $z_{i}\leq 1$ and $z_{i-1}\leq 2z_{c}-1$.
Presuming now that $i-1$ topples, all but possibly one
particle will stick to $i$, giving $z_{i-1}\leq 0$.
Even if $i-2$ now topples, no particles can slide
past $i-1$ and so now $z_{i-2}\leq 0$. From now on
this is all that happens, toppled particles neither
slide nor cause any further topples.
The avalanche will eventually end with at most $z_{c}+1$
particles from the leading edge of the avalanche
moving beyond site $i$ and potentially contributing
to the dropsize.
More qualitatively, toppling is minimal
for $z_{in}\leq z_{c}$ and the bulk of the sliding is
limited to within the avalanche for
$z_{in}\geq z_{c}$, so for $z_{in}=z_{c}$ we should
expect particle
transport to be low. Less commonly considered measures
of avalanche size, such as the total number of topples
involved, must be used in this case.

\begin{figure} [h]
  \begin{center}
    \leavevmode
      \psfig{figure=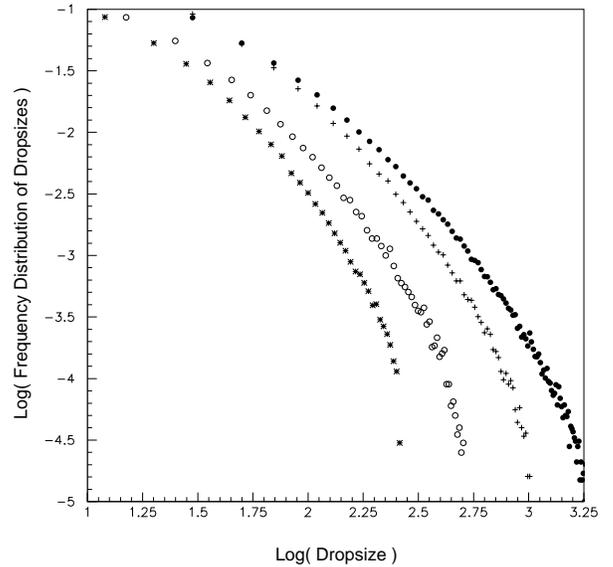,clip=true,width=8cm}
  \end{center}
\caption{Dropsize distributions for $z_{in}=1$ and $z_{c}=2$,
with varying system size.
Asterisks correspond to L=128, open circles to L=256,
crosses  to L=512 and filled circles to L=1024.}
\label{f:zin1}
\end{figure}

\begin{figure} [h]
  \begin{center}
    \leavevmode
      \psfig{figure=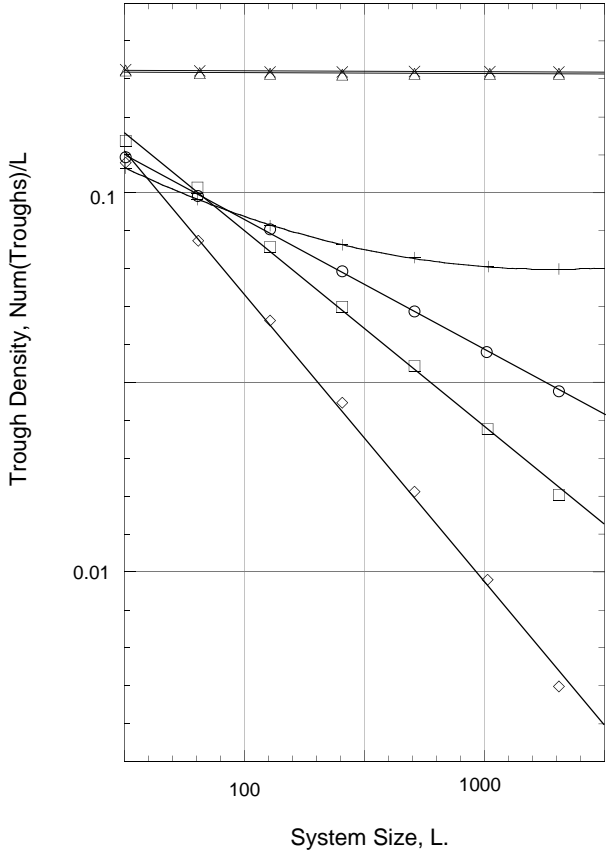,clip=true,width=8cm}
  \end{center}
\caption{Trough densities for $z_{c}=2$ against system size.
Circles refer to $z_{in}=4$, squares to $z_{in}=3$,
diamonds to $z_{in}=2$, plus-signs to $z_{in}=1$,
triangles to $z_{in}=0$ and crosses to $z_{in}=-\infty$.}
\label{f:troughs}
\end{figure}

For $z_{in}=1$ and $z_{c}=2$, the dropsize distributions
broaden with the system size, as shown in fig.\ref{f:zin1}.
There is perhaps some hint of a lessening in the broadening
for the larger systems. However, long processing times
have resulted in statistically
noisy data which it is
difficult to analyse. An alternative does exist, since
when $z_{in}\leq z_{c}$ the avalanche size is
bounded above by the distance between troughs,
so a system with a finite concentration of troughs must
have a finite correlation length
and hence cannot display SOC.
The trough density, being just a single value, is much less
susceptible to noise and can be
measured reliably for larger $L$.
The distribution of trough densities for $z_{c}=2$
and for $L$ up to 2048 is given in fig.\ref{f:troughs}.
That $z_{in}=1$ is not SOC is now evident as the trough
density tends to a finite value, although this value
is small, corresponding to a large correlation length,
which explains the observed broadening of the dropsize
distribution in small systems.
For $z_{in}=z_{c}=2$, the trough density approaches zero
with system size faster than for any other value
of $z_{in}$, corresponding to a singular correlation length
and SOC behaviour.
These results are borne out by the mean-field analysis
in section~\ref{sec:MFT}.

Another useful measure to consider is the average slope $S$,

\begin{equation}
S=\frac{1}{L}\sum_{i=1}^{L} z_{i},
\end{equation}

\begin{figure} [h]
  \begin{center}
    \leavevmode
      \psfig{figure=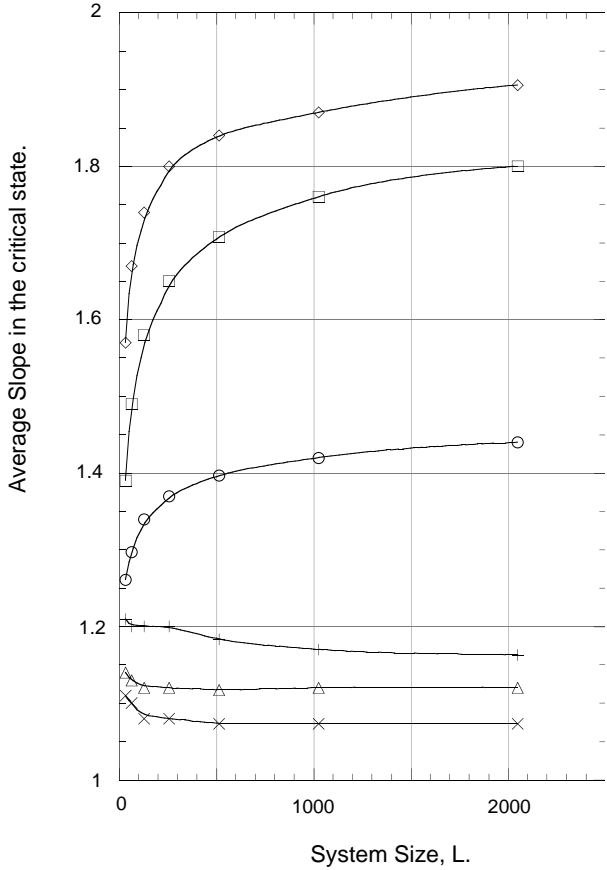,clip=true,width=8cm}
  \end{center}
\caption{Average slopes for $z_{c}=2$ against system size.
Circles refer to $z_{in}=4$, squares to $z_{in}=3$,
diamonds to $z_{in}=2$, plus-signs to $z_{in}=1$,
triangles to $z_{in}=0$ and crosses to $z_{in}=-\infty$.}
\label{f:slopes}
\end{figure}

\noindent{where $\tan^{-1} (S)$ is
the {\em angle of repose}.
The variation of $S$ with $L$ and $z_{in}$ is
given in fig.\ref{f:slopes}. For $z_{in}<2$ the lines converge
on values in the range $1<S<1.2$.
It is already known~\cite{Feigenbaum}
that for $z_{in}=4$ the
slope for $L\rightarrow\infty$ is $S=\frac{3}{2}$.
Comparing this line to those for other $z_{in}$
in fig.\ref{f:slopes},
it could be judged that the curves for
$z_{in}=2$ and $z_{in}=3$ are both tending to $S=2$
in a similar manner. Since convergence is slow, however,
verifying these asymptotic
limits is impossible from this data alone.
Taking logarithmic plots to project the lines
further, as was done successfully for the standard
model~\cite{Feigenbaum}, fails here as it gives reasonable
straight line fits for a range of limiting slopes.
Instead, a rough prediction of the asymptotic slope
for $z_{in}=2$ is given by the following qualitative
argument.
}

For $z_{in}\leq z_{c}$ particles leaving the pile
cannot cause the rightmost site $i=L$ to topple,
although if $z_{L}<z_{in}$
they will stick there and increase $z_{L}$.
Once $z_{L}$ is in the range
\mbox{$z_{in}\leq z_{L}\leq z_{c}$},
all particles will slide over $L$,
so $z_{L}$ is fixed until the
end of the current avalanche.
The same thing will then happen for site $L-1$, then $L-2$,
and so on. Thus, the effect of any avalanche reaching the
right-hand boundary is to leave behind a cluster of sites
with slopes in the interval $[z_{in},z_{c}]$.
Furthermore, the slopes in this region are stable
under subsequent particle addition
on any site to its left.
Similar clusters should grow throughout the pile
for similar reasons, although their
right-hand edges will not be fixed.
This has special significance for $z_{in}=z_{c}$,
since the clusters of sites will all have
slope $z_{c}$ and $S$ is
expected to tend to a value at least close to $z_{c}$.
Given that this system becomes dominated by sites
of critical slope, the $z_{c}+1$ particles
that slide from any avalanche should all leave
the pile, thus the modal dropsize should
also be the maximum one, $z_{c}+1$.
This has been confirmed by the numerical studies.

\section{Mean-Field Analysis}
\label{sec:MFT}

Christensen and Olami~\cite{RNN} successfully formulated
a random nearest neighbour
mean-field model for critical height sandpiles.
Directly extending their procedure to critical slope
models has proved to be intractable, but the principle
of taking random neighbours
in an infinite system can still be applied to the
MILLS model with some success.
Utilising this approach is non-trivial, however,
as the dynamics of the
system change significantly as $z_{in}$ is varied.
In this section we present a range of models
for $z_{in}\leq 4$ with $z_{c}=2$, comparing their
predictions to the numerical findings.
All these models are cast in terms of
the slope distribution $\{S_{n}(t)\}$,
where each $S_{n}(t)$ is defined to be the
proportion of sites in the system at time $t$
with slope $n$.
For convenience, we also define

\begin{equation}
S_{T}(t)=\sum_{i=-\infty}^{0}S_{i}(t),
\end{equation}

\noindent{which is the trough density.
As these slopes represent the sandpile in its stable
state,
the maximum value any $z_{i}$ can take is two and
}

\begin{equation}
\label{add}
S_{T}+S_{1}+S_{2}=1.
\end{equation}

The simplest case to analyse is $z_{in}=-\infty$, where
any avalanche propagates back to the first trough
and all toppled particles are
removed from the system. Particle addition alone
has the effect of increasing the slope of
a randomly chosen site from $n$ to $n+1$,

\begin{eqnarray}
\label{deci}
S_{n} & \rightarrow & S_{n} - \frac{1}{L}, \\
\label{incip1}
S_{n+1} & \rightarrow & S_{n+1} + \frac{1}{L},
\end{eqnarray}

\noindent{and in a similar manner
decreasing the slope of another randomly
chosen site by one. Troughs are
assumed to be primarily sites of slope zero,
so increasing the slope of a trough
will give a site of slope one, whereas the effects
of decreasing the slopes of troughs are ignored.}

When a particle is added to a site of slope two,
(\ref{deci}--\ref{incip1}) must be replaced
by equivalent rules for the whole of the subsequent
sequence of topples.
There are two cases to consider~-
if the site to the immediate left
(here chosen at random) has slope one,
a slide event occurs, removing
two particles from the system but leaving
$\{S_{n}(t)\}$ unchanged. Otherwise, we get a full
avalanche which only alters the slopes at its leading
edge (where the particle was added)
and its trailing edge (at the trough that halted the
toppling). The net result of both these edge effects
is the lose a trough and a site of slope two,
and to gain two sites of slope one.
It is now straightforward
to write down rate equations for $\{S_{n}(t)\}$,
with the timescale normalised to $L$ particle additions
per unit $t$,

\begin{eqnarray}
\label{inftys2}
\frac{dS_{2}}{dt} & = & -2S_{2} + S_{1} + S_{1}S_{2}, \\
\label{inftys1}
\frac{dS_{1}}{dt} & = & -2S_{1} + S_{T} + 3S_{2} - S_{1}S_{2},
\\
\label{inftyst}
\frac{dS_{T}}{dt} & = & -S_{T} + S_{1} - S_{2}.
\end{eqnarray}

These equations can be justified directly - for instance,
the first term on the right hand side of~(\ref{inftys1})
corresponds
to particle addition onto or adjacent to a site of slope one,
the second term is for a particle added to a trough,
the third term accounts for particle addition
adjacent to or onto a site of slope two,
where the latter instance invokes an avalanche,
and finally the
fourth term subtracts out trivial slide events.
In practice, however, it is simpler to derive
these equations by writing
down all nine possible combinations of pairs of sites,
ie. (TT) (T1) (T2) (1T) (11) (12) (2T) (21) and (22),
then computing directly the effects of particle addition
to the right hand site and weighting them accordingly.

The equations~(\ref{inftys2}--\ref{inftyst}) converge
on the stable solution
$(S_{T},S_{1},S_{2})=(\frac{1}{6},\frac{1}{2},\frac{1}{3})$,
in good agreement with the numerical values (0.18,0.48,0.34).
Since the influence of sites with strictly negative
slope have been ignored, this mean-field theory
actually covers the range of all $z_{in}\leq 0$, and
indeed the agreement still holds up to $z_{in}=0$.
For $z_{in}=1$ the avalanching
particles can no longer be ignored and new terms
need to be added to account for where they come to rest.
The expected number of particles brought into motion during an
avalanche depends on the state of the site to the immediate
left
of the site where the first particle was added.
If this site has slope less than two then just
the one site topples. Otherwise, the number of sites
that topple is half the average distance between
troughs. In this random neighbour model, the
probability that two nearest troughs are a distance
$r$ apart is $(1-S_{T})^{r}S_{T}$.
Consequently, half the mean distance between troughs is
$(1-S_{T})/2S_{T}$ and the total number of sites
that topple during an avalanche is on average
$S_{T}+S_{1}+S_{2}\left[ (1-S_{T})/2S_{T}\right]$.
Each toppled site releases two particles, so we
will need a factor of two. Furthermore, since
avalanches are initiated by particle addition to sites
of slope two, we will also need a factor of $S_{2}$.
Thus, the expected number of toppling particles
per unit time is

\begin{equation}
\lambda(t) = S_{2}\left( 2(1-S_{2})+
\frac{1-S_{T}}{S_{T}}S_{2}\right) .
\end{equation}

\noindent{Each of these $\lambda(t)$ particles
will stop on a trough, creating a site with
slope one. This will also cause the site to the immediate
left to decrease its slope by one. Moreover, we know that
the particle passed over this site, so it
cannot be a trough. Therefore it must have slope
one or two with probability $S_{1}/(1-S_{T})$ or
$S_{2}/(1-S_{T})$, respectively.
Since we are ignoring any correlations between where
a particle starts moving and where it stops,
all we need to do is to add
extra terms to the right hand sides of
(\ref{inftys2}--\ref{inftyst}) to account for the changes
in the slopes caused by the $\lambda(t)$ particles coming
to rest,
}

\begin{eqnarray}
\frac{dS_{2}}{dt}&=&-2S_{2}+S_{1}+S_{1}S_{2}
-\lambda\left( \frac{S_{2}}{1-S_{T}}\right) , \\
\frac{dS_{1}}{dt} & = &-2S_{1}+S_{T}+3S_{2}-S_{1}S_{2}
+\lambda\left( 1+\frac{S_{2}-S_{1}}{1-S_{T}}\right) , \\
\frac{dS_{T}}{dt} & = &-S_{T}+S_{1}-S_{2}
+\lambda\left( \frac{S_{1}}{1-S_{T}}-1\right) .
\end{eqnarray}

\noindent{These equations now converge to
$(S_{T},S_{1},S_{2})\approx (0.09,0.65,0.26)$,
whereas numerically the slope distribution is
$(0.07,0.68,0.25)$.
Note that the trough density $S_{T}$ is small
but non-zero, corresponding
to a large but finite correlation length.
Thus, as noted in the previous section,
large system sizes are required to demonstrate
that this system is not SOC.
}

For $z_{in}=2$, the unique nature of the avalanche
requires a new set of rate equations, which can be
derived in a similar manner to before
by considering the consequences of
particle addition to a pair of sites
with arbitrary slopes,

\begin{eqnarray}
\frac{dS_{2}}{dt}&=&S_{T}(S_{1}-2S_{2})+S_{1}^{2}
+\mu\left( \frac{S_{1}}{1-S_{2}}-1\right) ,\\
\frac{dS_{1}}{dt}&=&-2S_{1}^{2}+S_{T}(3S_{2}-S_{1}+S_{T})
+\mu\left( 1+\frac{S_{T}-S_{1}}{1-S_{2}}\right) ,\\
\frac{dS_{T}}{dt}&=&-S_{T}+S_{1}-S_{2}S_{1}
-\mu\left( \frac{S_{T}}{1-S_{2}}\right) ,
\end{eqnarray}

\noindent{where $\mu (t)=S_{2}(2+S_{2})$ is
the expected number of particles
that emerge to the right of the particle addition.
Although these equations admit the steady solution
$(S_{T},S_{1},S_{2})\rightarrow(0,0,1)$
with $S_{1}\sim 1-S_{2}$ and $S_{T}\sim 0$,
cubic corrections are required for it to be stable.
Numerically, the slopes are $(0.01,0.08,0.91)$ for
$L=2048$ but, as mentioned in section~\ref{sec:results},
the asymptotic slope appears to be $S=2$ which demands
the slope distribution $(0,0,1)$.
}

For $z_{in}=3$, all the sites involved in an
avalanche resolve in a highly non-trivial way,
so calculating rates
between the states before and after an avalanche
becomes impossible.
Instead, we introduce the
following {\em local slope argument} to
predict the slope distribution in the critical state.
Within an avalanche, a toppled site deposits
a pair of particles to its right, which, after any
inertial motion, settle and possibly cause
further toppling.
Keeping our random neighbour approach, the expected
number of further topples caused takes
the same constant value $E_{t}$ throughout
the avalanche. If $E_{t}<1$ then
avalanches would die out exponentially quickly,
particles will not get transported from the pile
and so the sandpile would build up, increasing the
average slope and hence also $E_{t}$.
If $E_{t}>1$, avalanches would grow exponentially
and large numbers of particles would leave
the pile, decreasing $E_{t}$.
Although transport from the pile
is not actually catered for by this model, we infer
that $E_{t}=1$ in the critical state,
so finding an expression for $E_{t}$ in terms
of $\{S_{n}(t=\infty)\}$ gives an additional
constraint which might help to fix
the slope distribution.

This local slope argument
has several deficiencies, apart from
the obvious drawbacks of the
random neighbour approach.
Back-avalanches have been
ignored, so this method will certainly
break down for $z_{in}\leq 2$.
The effects of single particle
addition has not been accounted for, which will 
further exasperate the situation when
$S_{2}$ is small.
Although this gives a limited range of applications
for this model, it has the advantage of
being conceptually and mathematically simple.
For instance, in the standard model $z_{in}=4$,
adding two particles to any site with a slope
greater than zero will cause one further topple, so
$E_{t}=1-S_{T}$.
Requiring that this equals one gives $S_{T}=0$,
as expected.

For $z_{in}=3$, consider the effect of adding the
particles to two adjacent sites, $i$ and $i+1$.
If $z_{i}<2$, then both particles would
stick to $i$, giving the site a slope
$z_{i}+2$, which will topple if $z_{i}=1$.
However, if $z_{i}=2$ then
the last particle added will slide onto the
adjacent site, resulting in
$z_{i+1}\rightarrow z_{i+1}+1$ and $z_{i}$ remaining
unchanged. Thus, the expected number of further topples
is

\begin{equation}
E_{t}=S_{1}+S_{2}^{2}.
\end{equation}

\noindent{Setting this equal to one and
employing~(\ref{add}), we get an equation
for the trough density,
}

\begin{equation}
\label{trough}
S_{T}=S_{2}(S_{2}-1).
\end{equation}

\noindent{By definition, all the $S_{i}$
must lie in the range [0,1]. However,~(\ref{trough})
is strictly negative for
$0<S_{2}<1$, so for consistency we must have
$S_{2}=0$ or $1$. Avalanches could never begin
in a system with $S_{2}=0$, so we conclude that
$(S_{T},S_{1},S_{2})=(0,0,1)$, in agreement
with the numerical work.
}

\section{Conclusions and Discussion}
\label{sec:summ}

It has been demonstrated that the MILLS model with $z_{c}=2$
displays self-organized criticality for systems with
$z_{in}\geq z_{c}$, but SOC is lost when $z_{in}<z_{c}$.
The SOC state for $z_{in}\geq z_{c}$ was identified by the
broadening of the dropsize distribution with the
system size, whereas for $z_{in}<z_{c}$ the existence of
a finite correlation length was found both numerically
and from the mean-field theories.
Preliminary numerical results for $z_{c}$ up to 4 demonstrate
that the transition point $z_{in}=z_{c}$ holds more generally.
We postulate that for all $z_{c}>1$ the model
is SOC only when $z_{in}\geq z_{c}$,
and now argue why this should be so.

An essential requirement of any SOC system is that
it is governed by local driving forces~\cite{BTW}.
For $z_{in}>z_{c}$, inertial effects only
come into play inside an avalanche, since
it is only within an avalanche that sites with slopes
greater than $z_{c}$ can occur. Thus, the couplings
remain short-range and SOC behaviour is preserved.
For $z_{in}<z_{c}$, however, particles can slide far
beyond the avalanche, so interactions become
non-local and the system ceases to be SOC.
We might na‹vely expect that a system with
$z_{in}=z_{c}$ will
also fail to be SOC, especially since the system
becomes dominated by sites of slope $z_{c}$, which
should allow particles to slide through large distances.
However, the exact analysis in section~\ref{sec:results}
proved that only a tiny proportion of particles
move beyond the avalanche. In fact, the vast
majority move by just one site,
which is certainly a local driving force,
so we can conclude that a system with $z_{in}=z_{c}$ is SOC.

In the experiments carried out on piles of
rice~\cite{nature}, power-law behaviour was only
observed for rice with sufficiently large aspect ratio.
In section~\ref{sec:model} it was argued
that $z_{in}/z_{c}$ should increase with the
particle aspect ratio. Thus, if the aspect ratio
is increased such that $z_{in}$ becomes greater
than $z_{c}$, the MILLS model gives a possible explanation
for the appearance of power-laws~- the system
has crossed over into the SOC regime.
However, many non-SOC systems also exhibit power-law
behaviour, and an alternative model has recently been
proposed by Newman and Sneppen~\cite{Newman}.
Here, they claim that an SOC approach is invalid
when the power-laws have characteristic exponents
in the vicinity of 2, as was observed for the rice piles.
Their model is instead driven by external noise, but
still predicts a crossover from power-law behaviour.

Regardless of its experimental applications, the MILLS
model represents an interesting addition to the range
of sandpile systems already studied, in that it
provides an inertial sandpile model which is SOC
for a broad band of parameter space - that is,
without the need for fine tuning - and independent
of the system size.
Extending this model to higher dimensions
or to critical height sandpiles should in principle be
straightforward, but since our original motivation
was to try and account for the results from 
the rice pile experiments, we chose
to base our model on that system.

\end{document}